# Proximity-induced superconductivity and quantum interference in topological crystalline insulator SnTe thin film devices

Robin-Pierre Klett[1], Joachim Schönle[2], Andreas Becker[1], Denis Dyck[1], Karsten Rott[1], Jan Haskenhoff[1], Jan Krieft[1], Torsten Hübner[1], Oliver Reimer[1], Chandra Shekhar[3], Jan-Michael Schmalhorst[1], Andreas Hütten[1], Claudia Felser[3], Wolfgang Wernsdorfer[2,4] & Günter Reiss[1]

[1]Center for Spinelectronic Materials & Devices, Physics Department, Bielefeld University, Universitätsstraße 25, 33615 Bielefeld, Germany

[2]CNRS, Institut NEEL and Université Grenoble Alpes, 38000 Grenoble, France

[3]Max Planck Institute for Chemical Physics of Solids, 01187 Dresden, Germany

[4]Karlsruhe Institute of Technology, Physics Department, Wolfgang-Gaede-Straße 1,

76131 Karlsruhe, Germany

## ABSTRACT

**Topological crystalline insulators represent a new state of matter, in which the electronic transport is governed by mirror-symmetry protected Dirac surface states. Due to the helical spin-polarization of these surface states, the proximity of topological crystalline matter to a nearby superconductor is predicted to induce unconventional superconductivity and thus to host Majorana physics. We report on the preparation and characterization of Nb-based superconducting quantum interference devices patterned on top of topological crystalline insulator SnTe thin films. The SnTe films show weak antilocalization and the weak links of the SQUID fully-gapped proximity induced superconductivity. Both properties give a coinciding coherence length of 120 nm. The SQUID oscillations induced by a magnetic field show $2\pi$ periodicity, possibly dominated by the bulk conductivity.**

# THE MANUSCRIPT

*Introduction*

In the past several years, tremendous efforts have been made in investigating new topological states of matter, and it has been widely recognized that the family of topological materials is highly diversified and widespread [1]. One of its species is formed by topological crystalline insulators (TCI), where topological protection is due to the symmetry of the crystal structures. The first predicted [2] and experimentally verified [3] class of TCI materials was found within the IV-VI semiconductors, with SnTe as a representative model material. SnTe crystallizes in rock-salt structure and the symmetry responsible for its topological nature is the reflection symmetry with respect to the (110) mirror planes [2,3]. The appearance of unconventional boundary modes in such topologically non-trivial phase has been detected experimentally via several techniques [4-6], manifesting themselves in linearly dispersing chiral topological surface states (TSS). One among the most mesmerizing consequences of the TSS and non-trivial topology arises, if the material itself undergoes a superconducting phase transition. In this case, theoretical models predict the occurrence of topological (mirror) superconductivity [7]. More specifically, the four Dirac cones in the TCI surface Brillouin-zone give rise to host Majorana fermion quartets [8]. For intrinsic superconducting TCI a couple of smoking gun experiments have been made in In-substituted $Sn_{1-x}In_xTe$ [9,10], which shows a superconducting phase transition at $T_c = 3.5 - 4.7\,K$ while maintaining its non-trivial band structure [11]. However, the issue of extrinsic (proximity-induced) superconductivity in undoped SnTe has not been tackled yet experimentally. In this case, the system can undergo a superconducting phase transition due to its proximity to a nearby conventional s-wave superconductor (SC) [12]. Here, we report on the fabrication and characterization of SC – TCI hybrid microstructures. We patterned superconducting quantum interference devices (SQUIDs) made from Nb thin films on co-sputtered SnTe thin films such that bridges from SnTe form weak links between superconducting Nb wires. The resistance of the SnTe shows a weak antilocalization in its response to magnetic field and the SnTe weak links become fully superconducting. We additionally studied the response of the SQUID ring to a DC magnetic field. If theoretically predicted Majorana physics tend to be present due to TSS attached to nearby s-wave SCs, the conventional ($2\pi$-periodic) SQUID relation

$$\phi_0 = \frac{h}{2e} = A_S \, \delta B \qquad (I.),$$

where $\phi_0$ is the magnetic flux quantum, $A_S$ the effective area of the SQUID and $\delta B$ the oscillation period of the critical current response, would be superimposed by an anomalous contribution, arising from $4\pi$-periodic transport characteristics. A SQUID interference pattern hosting $4\pi$-periodic physics would be an ultimate proof of topological superconductivity [13] and hence, the presence of Majorana fermions. Our measured data show an absence of any unconventional contribution and thus follows the behaviour of conventional SQUIDs.

## *Sample preparation*

In this section, we describe the growth procedure of the TCI – SC hybrids and give a detailed description of the lithography and patterning process. First, thin films of SnTe are grown on MgO substrates at 150°C via co-sputtering of Sn and Te. We chose insulating MgO (001) substrates due to their cubic crystal structure and their good lattice match to the SnTe lattice constant $a_{SnTe} = 0.63$ nm $\approx \sqrt{2}\, a_{MgO}$ [14]. The base pressure of the sputter system was $5 \cdot 10^{-10}$ mbar, the Ar-pressure was p ~ $2.5 \cdot 10^{-3}$ mbar. As shown in Fig. 1(a), x-ray diffraction (XRD) in $\Theta$-$2\Theta$ geometry clearly produces only the (002) and (004) diffraction peaks. Thus, under these conditions 40 nm SnTe thin films grow polycrystalline (PC) with strong (001) texture. In the next step the TCI thin films are coated with Nb, a well characterized and established s-wave superconductor. The Nb was deposited without vacuum break to prevent TCI surface contamination and to allow a high transparency at the SnTe/Nb interface [15, 16]. The nominally deposited 27.5 nm Nb films grow in (110) texture with low strain on top of the SnTe rock-salt lattice, which is confirmed with the XRD peak at 38.3° in Fig 1(a). The x-ray reflection data presented in Fig. 1(b) verify smooth film growth with a fitted roughness of $\sigma \approx 0.1$ nm. A low surface roughness has recently been identified as a crucial precondition for the presence of TSS in TCI films [17,18]. Since Nb is known to be a chemically very active metal, the film quality (and the superconducting properties) often suffer from lithographic and etching processes during device patterning. We countered this issue by capping the TCI/SC hybrid with 2.5 nm Ta and 5 nm Ru as protection for subsequent fabrication steps. The SQUIDs were patterned with conventional e-beam lithography. After resist development, the film stacks are patterned by Ar-ion dry etching. Controlled via secondary ion mass spectroscopy, we stopped the sample etching once the Nb was removed. The patterned SQUIDs (exemplarily shown in

Fig. 2) have an effective area A=16 µm² and the Nb lines have a width of w = 200 nm. The length L of the two Josephson junctions was varied between L = 50 – 200 nm. The samples were characterised electrically both by dc measurements of the film resistance as well as by a current-biased lock-in technique for differential resistance dV/dI scans.

## Results

*SnTe thin films – Magnetoresistance*

Before starting the characterization of the proximity induced superconductivity and the SQUID data, we investigated the transport properties of bare SnTe thin films using patterned Hall bar microstructures as schematically shown in the inset of Fig. 3(a). The determination of the Hall coefficient with Hall measurements at 2 K (not presented) show that the charge-carrier type is strongly hole-like and the carrier density is about n ≈ 4.5·$10^{20}$ cm$^{-3}$. As SnTe is known to tend intrinsically to high carrier concentrations due to Sn vacancies (and hence to p-type behaviour), this is consistent with current reports of other researchers [19]. An established all-electrical method of testing the presence of surface state transport are measurements of the longitudinal magnetoresistance (MR), where a weak antilocalization (WAL) is expected due to surface states. The MR-data shown in Fig. 2(a) are plotted as relative change in magneto-conductivity (MC) Δσ$_{2D}$ = σ(B) – σ(0) and show sharp cusp-like MR, which can indeed be attributed to WAL and evaluated with the Hikami-Larkin-Nagaoka (HLN) formalism

$$\Delta\sigma_{2D} = \alpha \frac{e^2}{2\pi^2\hbar^2} \left( \ln\left(\frac{\hbar}{4eBL_\phi^2}\right) - \psi\left(\frac{1}{2} + \frac{\hbar}{4eBL_\phi^2}\right) \right) \quad (II.),$$

where e is the electron charge, ℏ is the Planck's constant, L$_\Phi$ is the phase-coherence length of a charge carrier in a given surface channel, ψ(x) is the digamma function, B is the out-of-plane applied magnetic field and α$_0$ a dimensionless transport parameter [20, 21]. Fitting the data yield information about L$_\Phi$ and α$_0$, which are plotted both as function of temperature T in Fig. 3(b). L$_\Phi$ is increasing steadily with decreasing T from L$_\Phi$ ≈ 20 nm at 20 K to L$_\Phi$ ≈ 120 nm at 2 K, which is the same order of magnitude as stated for other topological material [21 - 23]. The theoretical prediction for the latter is supposed to yield α$_0$ = -0.5 for one TSS contributing to transport [20]. As shown in Fig 3(b), the fit of our data result in α ≈ - 0.5 over the entire temperature range demonstrating topological protected transport. Anyhow, in TCI material four TSS exist on the surface Brillouin zone (SBZ) each entering with an additional α$_0$ = -0.5 contribution. If one considers that TSS occur at the top and bottom interface, one would end up with a sum of |α| = 4 [24]. A plausible explanation for finding only the contribution of one TSS per surface is most likely due to the valley degeneracy of the SnTe SBZ, giving rise to two different coupling scenarios: *intra-* and *inter-surface valley coupling*. The first effect can be observed, when a carrier is able to scatter coherently between Dirac valleys located on the same surface. Accordingly, a long coherence length results in strong intra-valley coupling, and thus a smaller α - similar behaviour is typically observed for other 2D Dirac valley materials like graphene [25]. The second scatter mechanism is *inter-surface valley coupling* between top and bottom surface valleys. In this coupling regime, charge carriers can scatter coherently between the top and bottom SBZ Dirac valleys via bulk. This scenario has been observed by several other groups [24, 26, 27] and was attributed to the high bulk carrier concentration. Since WAL, however, is predominantly a 2D phenomenon, the bulk bands are unlikely to be the origin of the WAL [21, 24, 28]. Thus, the WAL features seen in Fig. 3(a) can be considered as consequence of spin-momentum locking of the TCI TSS, but the bulk bands have strong influence on the WAL due to coupling to TSS transport channels.

*SnTe/Nb SQUIDs – proximity induced superconductivity*

In Figs. 4(a) – (c), results for the differential resistance (dV/dI) as a function of the bias current $I_{sd}$ for Josephson junction lengths ($L_J$) of 50 nm, 120 nm and 200 nm are presented, respectively. If not further stated, all measurements were taken at 85 mK. The lengths of the weal links cover the range of the phase-coherence length $L_\Phi$ of the surface state electrons evaluated from the WAL data by HLN - analysis. Hence, if surface state properties are contributing to the transport, the features resulting from proximity of the TSS to the Nb should be most prominent for $L_J$=50nm in the dV/dI data (Fig. 4(a)). The individual sweeps can be separated into two sections: first, when $I_{sd}$ is lower than the critical current $I_c = 1.89\,\mu A$, a dissipationless supercurrent can flow through the system and the device exhibits a clear zero-resistant state. No spectroscopic features are observed in the gap, so that we can exclude, e.g. quasiparticle hybridization effects, which are known to appear in topological material. Second, for a bias current larger than $I_c$, the SQUID can be characterized by its normal state resistance $R_n \approx 12.29\,\Omega$. Small hysteretic behaviour of the switching between both regions is present. While sweeping back (forth) from high positive (negative) bias currents, one can identify small re-trapping currents $I_R$=1.39 μA. For these values, the dimensionless McCumber parameter $\beta = \frac{2eI_C R_n^2 C_g}{\hbar} \ll 1$, with $C_g = 1,42\,aF$ as geometrical capacitance of each Josephson junction (which is itself too low to yield in underdamping-like response), suggests that the SQUID is overdamped and no hysteresis is expected. Yet, a small hysteresis can be observed and a plausible explanation are self-heating effects [29]. The most prominent features in this dV/dI data are the strong peaks appearing at the outer-gap sides, while sweeping forth (back) from the resistive state into the superconducting gap at I = +/- 1.91 µA. Interestingly, when sweeping forth (back)from the inner non-resistive gap-region, the system jumps at the same current back into the resistive state, where the back (forth) biased current reveals the most peak. We attribute the peaks to the contribution of the TSS carrying supercurrents. In this picture, surface Andreev bound states (SABS) generated in analogy to the mirror-protected Dirac points on the (001) surface of the SnTe crystal can host Cooper pairs, which are more robust and manifest themselves as additional peaks outside the bulk-superconducting gap. *Hashimoto et al.* predicted such mirror-protected SABS hosted in superconducting TCI on the (001) surface, if an odd-parity potential is realized [30].

In Fig. 4(b) the length of the weak corresponds the estimated coherence length of the surface channels ($L_J = 120$ nm $\approx L_\Phi$), and hence resulting features from SABS are expected to be weakened, but still be present. The induced superconductivity still shows a zero-resistive behaviour in the gap, but the observed gap-flanking SABS peaks vanished. Interestingly, the superconducting gap flanks are not showing a sharp jump as in Fig. 4(a), but appear to be tilted outwards, which suggests that pure s-wave type is unlikely. If superconductivity is carried to some part by TSS, it is theoretically predicted that a mixture of s- and p-wave pairing is present [31]. A thermally induced hysteresis is still present, but strongly supressed compared to Fig. 4(a), which can be attributed to the lower critical current $I_c = 0.78$ µA (and a re-trapping current $I_R = 0.69$ µA) of the device. In Fig. 4(c) the differential resistance sweeps of a device with a weak link of $L_J = 200$ nm ($> L_\Phi$) is shown. For this length, an induced superconducting gap is still observable, but it does no longer exhibit a zero-resistance state, keeping a finite in-gap resistance of $R \approx 55.5$ Ω. Results for other samples suggest, that a zero resistance states exists for $L_J < 175$ nm. This result supports that the phase-coherence length of the surface currents and the coherence length of the Cooper-pairs are correlated and have crucial influence on the proximity-induced superconductivity in our films.

*SnTe/Nb SQUID oscillations*

The response of the differential resistance dV/dI to an out-of-plane magnetic field $B_z$ reveals clear oscillations in $I_c$ shown in the colour-code plot Fig. 5 for a length $L_J = 120$ nm and a field range from 0 to 1.1 mT (note, that the oscillations persist up to $B_z > 10$ mT). The blue regions correspond a SQUID resistance $R = 0\,\Omega$, while white and red areas represent finite resistance states. This mapping allows evaluating the critical current response for each value of the magnetic field $B_z$. The oscillations are periodic for $\delta B = 121{,}7\,\mu T$. If one takes a closer look at relation (I.), this corresponds to an effective area $A_S = 16.99\,\mu m^2$, which is in reasonable agreement with the dimensions of the SQUID rings characterized in our experiment, if one considers the London penetration length of Nb $\lambda_{Nb} = 350$ nm [32] on all sides of the square. Thus, we conclude the data show only conventional critical current response through the surface supercurrents, and hence no $4\pi$-periodic modulation is observed. Our finding agrees with the results of similar experiments performed by other groups on other topological materials, such as strained HgTe or $Bi_2Te_3$, $Bi_2Se_3$ compounds [32 - 34]. The reason for the absence of $4\pi$-periodicity may be a $2\pi$-signal poisoning arising from the huge amounts of supercurrents carried by bulk channels, which dominate the transport. This $2\pi$ channels coexist with a contribution of $k_y \neq 0$ ($k_y$ is the wave vector-component transversal to the moving direction $k_x$). Under these circumstances the predicted zero-energy Andreev bound excitations (Majorana fermions) are gapped out [35]. The small influence of the unique $4\pi$-periodic topological modes would than elusively vanish within this parasitic background. Here, we emphasize that (locally) gated measurements [36], as well as RF-transport measurements [37] are necessary next steps to address this problem and access new field of Majorana physics in TCI thin film systems.

## *Conclusion*

In conclusion, we have demonstrated WAL in the TCI SnTe and the first experimental evidence of fully-gapped superconductivity in SnTe/Nb heterostructures at $T = 85$ mK and a length of Josephson junction weak links below $L_J < 175$ nm. The phase-coherence length of the TCI extracted from these two properties show reasonable agreement. We additionally investigated the response of Nb-SQUIDs patterned on top of the SnTe to perpendicular magnetic field. The measured SQUID oscillations follow the relation describing a conventional SQUID interference pattern. Here, our results form a foundation for future investigations of proximity-induced topological superconductivity in the class of topological crystalline matter.

*Acknowledgements*

The financial support by the Deutsche Forschungsgemeinschaft (DFG, German Research Foundation) within the priority program SPP 1666 "Topological Insulators" is gratefully acknowledged.

*Author contributions*

All authors contributed equally to this work.

*Additional information*

Correspondence and requests for materials should be addressed to R.P.K. (rklett@physik.uni-bielefeld.de).

*Competing financial interests*

The authors declare no competing financial interest.

*Figures*

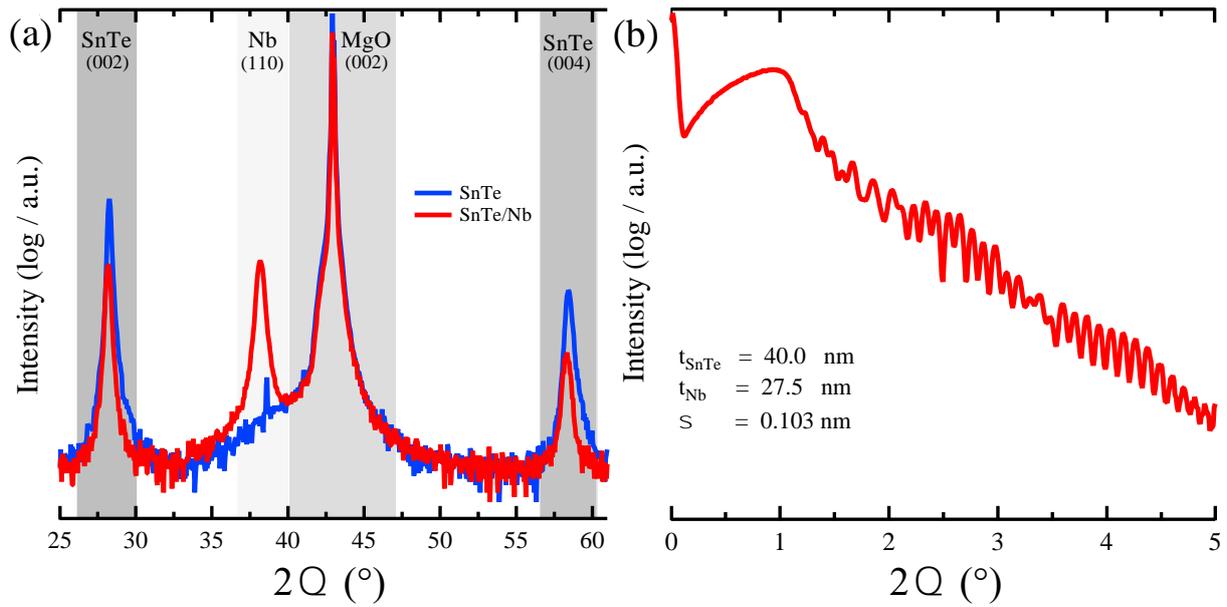

**Fig 1** (a) XRD scans taken for a 50 nm and a 40nm thick SnTe - 27.5 nm Nb bilayer grown on a MgO (001) substrate. The SnTe films show a polycrystalline growth with strong (001) texture. The subsequent Nb layer grows in (110) on top of the SnTe. (b) X-Ray reflection measurements of the bilayer structures proof the flatness of the films with a roughness of only 0.1 nm and indicate low strain at the interface of both layers.

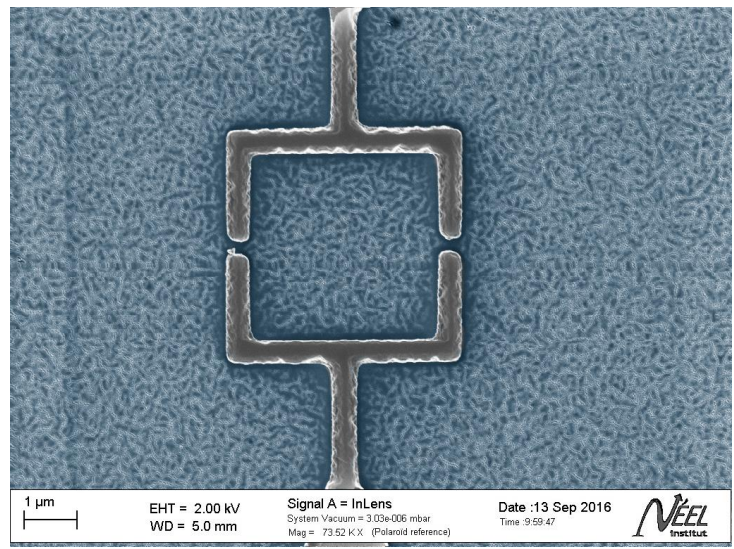

**Fig 2** False-colour SEM image of one of the fabricated SQUID samples. The Nb lines are highlighted in grey scale, while the SnTe film is pictured in dark blue. The nominally patterned inner area of the device is A=16 µm² and the length of the weak links is $L_J$ = 120 nm.

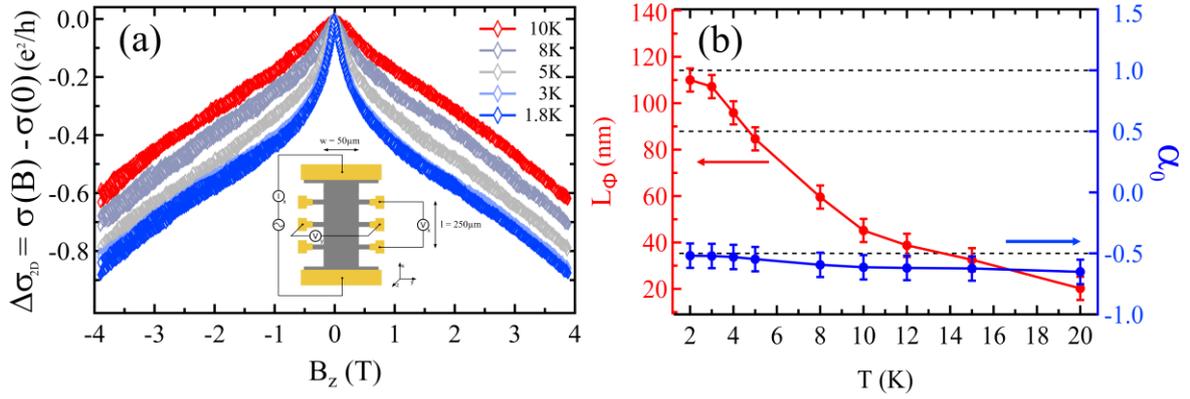

**Fig 3** (a) Results for the magneto-conductivity for several temperatures T against as a function of out-of-plane magnetic field $B_z$. The image in the inset shows the dimension of the micro-patterned Hallbar devices. The data indicate weak anti-localization and thus give evidence for two-dimensional electronic transport. (b) Fitting the data for small magnetic fields ($B_z < 0.5\,\text{T}$) with the HLN-formalism yield the phase-coherence length $L_\Phi$ and transport parameter $\alpha_0$, which are plotted in dependence of the temperature.

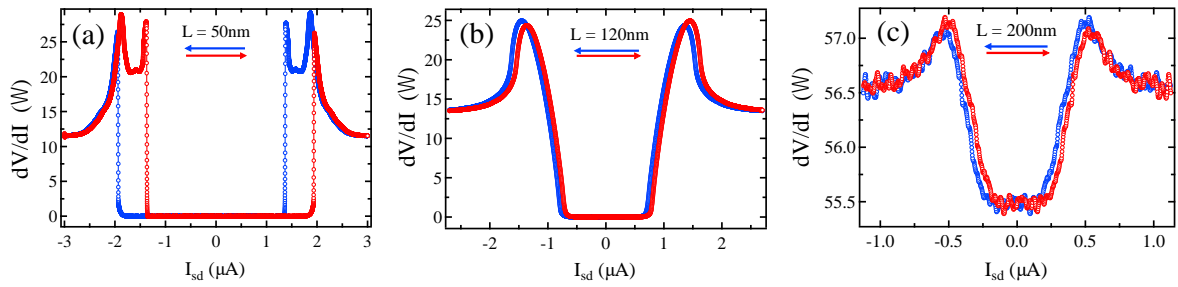

**Fig 4** Current-biased dV/dI sweeps for different length of the weak links as a function of the bias current $I_{sd}$. (a) For a weak link with $L < L_\Phi$ pronouced peaks in the differential resistance at the flanks of the induced superconducting gaps ($dV/dI = 0\,\Omega$) are observed. (b) For $L \approx L_\Phi$ the superconductivity is still fully gapped, but no flanking peaks occur. (c) For $L > L_\Phi$ the proximity superconductivity does not cover the entire weak link and the junctions remains resistive. The red / blue arrows indicate the sweep direction of the current.

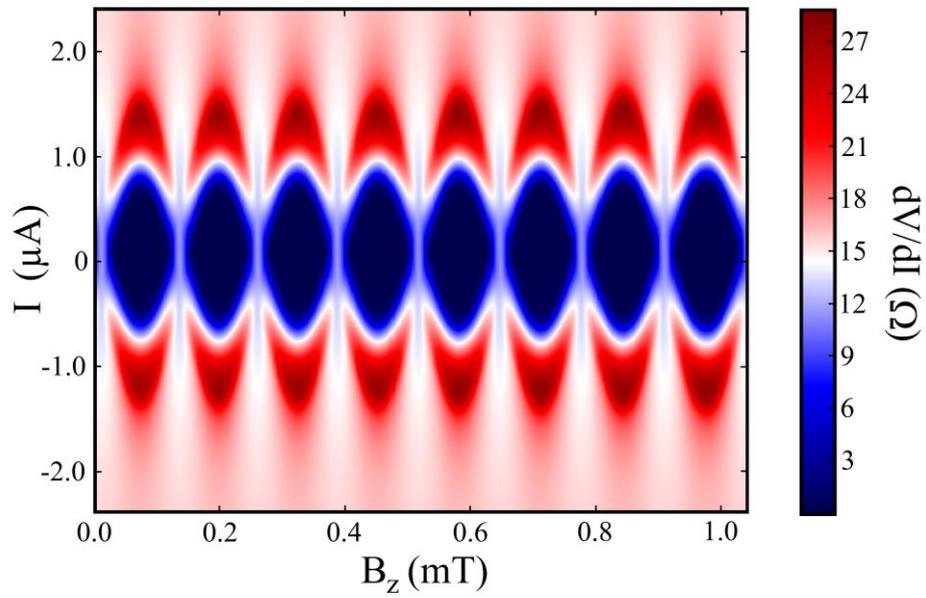

**Fig 5** dV/dI colour-plot showing the SQUID response (dV/dI as a function of $B_z$ and I) to an external out-of-plane magnetic field $B_z$ and bias current I. Clear oscillations are observable. The critical current $I_c$ oscillates with the period δB=121,7 µT that corresponds to an effective SQUID area of $A_S$ = 16,99 µm², obeying the conventional $\phi_0$-relation.